# Constructive Reuse of Free-Surface Ghost Wavefields in Marine Seismic Data: A Propagation-Based Framework

Hitoshi Mikada[1,2]

[1] Arc Geo Support Co., Ltd., Osaka, Japan

[2] Professor Emeritus, Kyoto University, Kyoto, Japan

## Abstract

Recent advances in marine seismic acquisition and processing have substantially improved the decomposition of free surface ghost wavefields. Building upon these developments, this study proposes a propagation-based framework in which decomposed ghost wavefields are treated as physically meaningful components that can be constructively reused as additional wavefield components. Starting from a finite inverse expansion of the ghost operator, a recursive formulation is derived and implemented through adaptive predictive decomposition. The formulation is subsequently extended to multidimensional wavefields through wavefield extrapolation and reciprocity, enabling decomposition into components associated with real and mirrored sources and receivers. The decomposed wavefields are then restored to a common acquisition geometry by survey sinking and combined through wavefield superposition, forming a preprocessing module that can be incorporated into conventional marine seismic processing workflows. Synthetic examples demonstrate increased signal strength through ghost reuse, spectral amplitude recovery of ghost-related notches including frequencies near DC, and clear decomposition of ghost-related wavefield components. Field-data examples acquired using a conventional shallow-towed airgun survey further demonstrate improvements in reflector continuity and coherent reflection energy, together with partial restoration of spectral amplitudes degraded by the free surface ghost, supporting the practical applicability of the proposed framework. The proposed framework therefore provides a processing-based complement to conventional ghost decomposition technologies and suggests a unified approach in which acquisition geometry and ghost processing are jointly optimised to maximise the usable wavefield information.

## 1. Introduction

Recent advances in wave-equation-based seismic imaging and inversion have substantially increased the demand for broadband seismic data. Modern wave-equation-based methods, including full-waveform inversion (FWI; Virieux and Operto, 2009), least-squares reverse-time migration (LSRTM; Dai and Schuster, 2013), and, more recently, Marchenko-based wavefield retrieval (Wapenaar et al., 2014), require accurate low- and high-frequency wavefield information for reliable subsurface imaging and inversion. Consequently, extending the usable bandwidth of seismic data has become one of the principal objectives of modern seismic imaging and inversion. Among various exploration environments, marine seismic surveys present a particularly challenging case because the usable bandwidth has long been limited by free surface ghosts.

Free-surface ghosts have long been recognised as one of the principal factors limiting the usable bandwidth of marine seismic data because they introduce characteristic spectral notches and attenuate low-frequency energy (Yilmaz, 2001; Carlson et al., 2007). Consequently, considerable effort has been devoted to extending the usable bandwidth through both acquisition and processing technologies.

Acquisition-based approaches have evolved through several complementary technologies for wavefield decomposition and broadband acquisition, including twin-streamer deghosting (Posthumus, 1993), PZ wavefield separation (Soubaras, 1996; Itenburg et al., 2006), over/under streamer acquisition (Moldoveanu et al., 2007), and variable-depth streamers (Soubaras and Dowle, 2010; Soubaras and Lafet, 2013). More recently, these developments have been further advanced by the introduction of ultra-low-frequency marine seismic sources (Dellinger et al., 2016; Ronen and Chelminski, 2017; Tellier et al., 2021).

Processing-based approaches include deterministic inverse filtering and predictive deconvolution (Claerbout, 1985; Robinson and Treitel, 1980), as well as wavefield-based methods such as surface-related multiple elimination (Verschuur et al., 1992; van Groenestijn and Verschuur, 2009). Whereas deterministic filtering seeks to compensate directly for the ghost response, wavefield-based approaches exploit the propagation properties of seismic wavefields to improve imaging and multiple suppression.

Although the constructive reuse of free surface ghost wavefields has not previously been investigated, wavefield components previously considered to be unwanted interference have been shown to be constructively reused to improve the accuracy of subsurface imaging. These approaches, known as Separated Wavefield Imaging (SWIM), have shown that selected wavefield components can be constructively utilised (Whitmore et al., 2010; Valenciano et al., 2014; Lu et al., 2015). These methods also raise an important question. Once free surface ghost wavefields have been successfully decomposed, should they simply be discarded, or should they instead be regarded as physically meaningful wavefield components that can be constructively reused?

Building upon these advances, the present study proposes a propagation-based framework in which free surface ghosts are reinterpreted as deterministic propagation phenomena rather than simply as unwanted wavefield components. A recursive formulation is derived from the finite inverse expansion of the ghost operator, providing the theoretical basis for recursive predictive decomposition. The formulation is subsequently extended to multidimensional wavefields through reciprocity, enabling source- and receiver-side ghost decomposition, survey sinking (Claerbout, 1985), wavefield superposition, and constructive reuse of ghost wavefields within a unified preprocessing workflow.

Starting from the finite inverse expansion of the ghost operator, the proposed framework is developed into a multidimensional wavefield propagation framework incorporating source- and receiver-side ghost decomposition, survey sinking, wavefield superposition, and the constructive reuse of decomposed ghost wavefields. The proposed methodology is validated using both synthetic and field-data examples acquired with a conventional shallow-towed marine seismic survey. The results demonstrate the practical feasibility of the proposed framework and its potential as a processing-based extension to conventional marine seismic acquisition for broadband imaging and inversion.

## 2. Theoretical formulation

### 2.1 Physical interpretation of free surface ghosts

Free-surface ghosts arise from the reflection of seismic wavefields at the sea surface, which acts as an approximately pressure-release boundary in marine environments (Yilmaz, 2001; Soubaras, 1996). As a result, seismic signals emitted from a source below the free surface and recorded by receivers towed at depth are accompanied by additional wavefield components that have been reflected at the free surface before being observed. These ghost components interfere with the primary wavefields and give rise to characteristic periodic notches in the frequency spectrum.

From a kinematic perspective, free surface ghosts can be represented as delayed replicas of the primary wavefields. For a source towed at a depth $d_s$ and a receiver towed at a depth $d_r$, the ghost contributions correspond to additional propagation paths involving reflection at the free surface. Under the assumption of vertical propagation beneath a locally planar free surface and a constant water sound velocity $V$, the associated time delays are given by $\tau_s = 2d_s/V$ for source ghosts and $\tau_r = 2d_r/V$ for receiver ghosts. Accordingly, the recorded seismic trace can be expressed as a superposition of the primary and ghost wavefields with characteristic time shifts determined by the acquisition geometry.

This representation highlights that free surface ghosts are not arbitrary noise components but coherent wavefield contributions governed by deterministic propagation paths and containing physically meaningful wavefield information. In the frequency domain, the superposition of primary and ghost wavefields modulates the observed wavefield spectrum, producing periodic spectral notches whose locations depend explicitly on the towing depths of the source and receivers. These notches limit the usable bandwidth and complicate subsequent seismic data processing. An equivalent mirror-image interpretation presented later in this paper provides the physical basis for the proposed propagation-based processing framework.

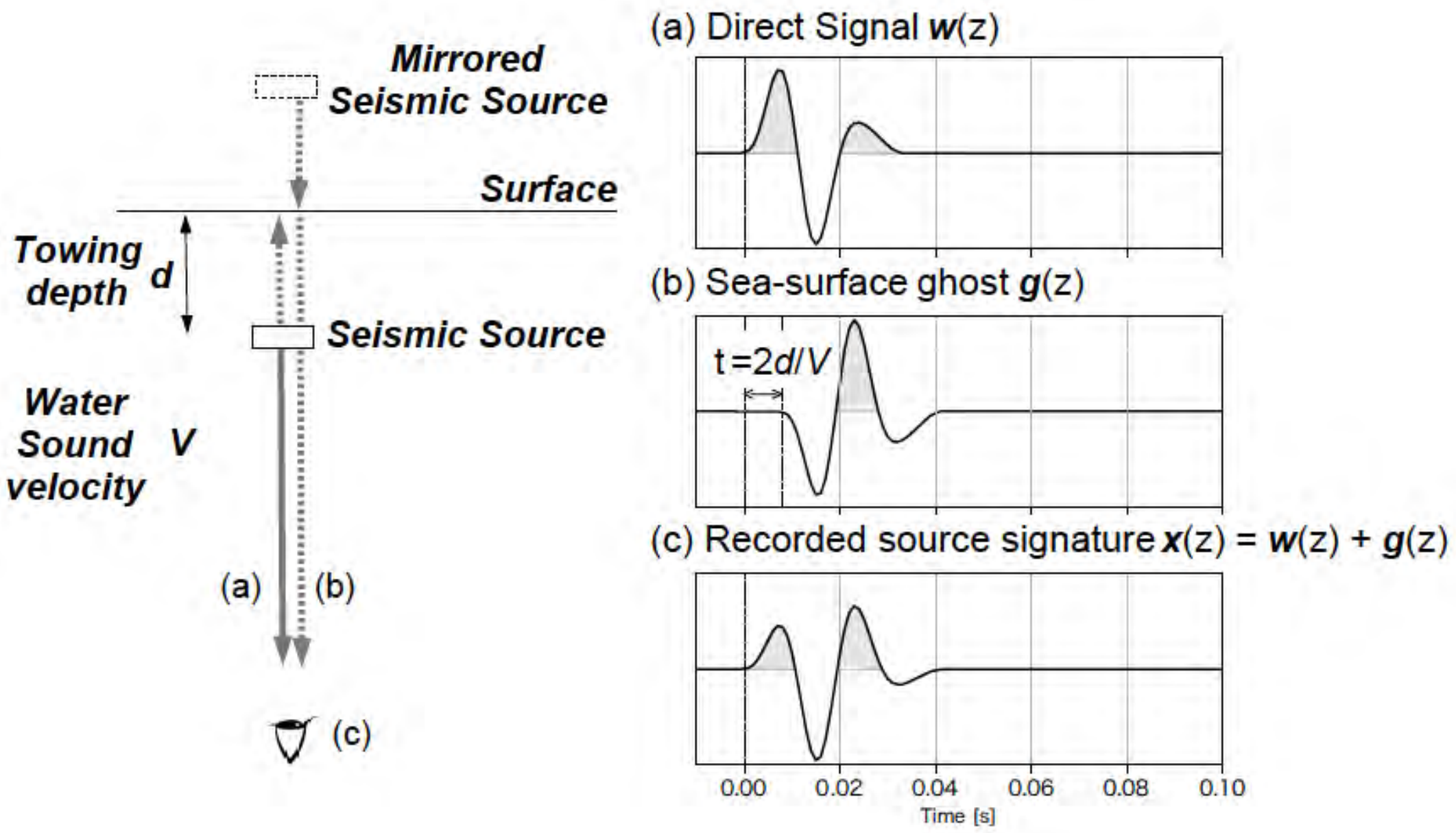

**Figure 1**. Conceptual illustration of the mirror-image interpretation of free surface ghosts. For simplicity, a 50-Hz minimum-phase Ricker wavelet is used to illustrate the deterministic relationship between primary and ghost components. In this figure, the source signal is emitted from a source placed at a depth of 6 m, corresponding to a source-ghost delay of 8 ms. Primary wavefields emitted from a seismic source and recorded by a receiver are accompanied by ghost wavefields reflected at the sea surface. These ghost components can be interpreted as observations generated by a virtual source located at mirror-image positions with respect to the free surface. The eye-shaped icon indicates the observation point where the source signature is recorded. By estimating the source wavelet, primary and ghost components can be temporally realigned according to their respective propagation paths and coherently summed, enabling constructive interference rather than spectral cancellation. Labels (a)–(c) correspond to the waveforms shown on the right.

## 2.2 One-dimensional ghost representation

Figure 1 illustrates the one-dimensional source-ghost model considered in this study. The receiver-side ghost is neglected for simplicity in this introductory step. The observed signature ($x(\zeta)$) is represented as the sum of the direct signature ($w(\zeta)$) and the ghost signature ($g(\zeta)$),

$$x(\zeta) = w(\zeta) + g(\zeta). \quad (1)$$

where $\zeta$ denotes the z-transform variable used in the one-dimensional formulation; the symbol $\zeta$ is adopted to distinguish it from the vertical spatial coordinate $z$ introduced later.

The ghost component is interpreted as a delayed replica of the direct signature generated by reflection at the free surface. Introducing the ghost operator ($E(\zeta)$), the ghost signature can be expressed as

$$g(\zeta) = -E(\zeta)w(\zeta). \tag{2}$$

Accordingly, the observed signature is expressed as

$$x(\zeta) = \big(I - E(\zeta)\big)w(\zeta). \tag{3}$$

This operator representation forms the basis for the finite inverse expansion developed in the following sections.

**2.3 Finite inverse expansion**

The coherent and deterministic nature of free surface ghosts suggests that they can be modelled explicitly and decomposed from the observed wavefield. In conventional marine seismic processing, the recorded seismic trace is commonly interpreted as the convolution of the source wavelet with a ghost operator representing free surface reflections at the source and receiver. In the frequency domain, this operator produces depth-dependent interference patterns that appear as periodic spectral notches.

Conventional approaches formulate ghost decomposition as an inverse-filtering problem. For a pressure-release free surface, the ghost operator is represented as the superposition of the primary wavefield and a delayed replica with opposite polarity. The corresponding inverse transfer function contains poles located on or close to the unit circle in the complex z-plane (Robinson and Treitel, 1980; Claerbout, 1985). Consequently, inverse filtering has generally been regarded as poorly convergent and susceptible to oscillatory behaviour.

In this study, the ghost operator is instead formulated in terms of a propagation operator acting on finite-length seismic records. Under this formulation, its inverse can be expressed as a finite expansion rather than an infinite recursive filter. This alternative representation forms the theoretical basis for the proposed wavefield decomposition and subsequent constructive reuse strategy of ghost wavefields.

For the one-dimensional model, the ghost operator may be expressed as the product of a free surface reflection coefficient and a propagation operator $P(\zeta)$,

$$E(\zeta) = rP(\zeta), \quad (4)$$

where r denotes the magnitude of the free surface reflection coefficient. For the ideal pressure-release free surface considered here, $r = 1$; in practical marine seismic applications, the magnitude of the reflection coefficient is commonly approximated as $r \approx 1$. The propagation operator reduces to a simple delay expressed as

$$P(\zeta) = \zeta^{-n_\tau},$$

where $n_\tau$denotes the propagation delay in samples, corresponding to the ghost delay $\tau = 2d/V$, with $d$ denoting the source depth and $V$ the water sound velocity. For the derivation below, $n_\tau$ is taken as an integer. Although the physical propagation time is not restricted to integer multiples of the sampling interval, the discrete representation adopted here is conceptually related to the propagation model of Goupillaud (1961), in which wave propagation is represented through successive discrete propagation steps.

Therefore,

$$P(\zeta)w(\zeta) = \zeta^{-\tau/\Delta t}w(\zeta),$$

Because the free surface acts as a pressure-release boundary, the ghost signature has opposite polarity and may therefore be written as

$$g(\zeta) = -rP(\zeta)w(\zeta).$$

The observed signature is therefore represented by

$$x(\zeta) = w(\zeta) + g(\zeta) = \big(I - rP(\zeta)\big)w(\zeta).$$

Accordingly, recovery of the direct signature formally requires application of the inverse operator,

$$w(\zeta) = \left(I - rP(\zeta)\right)^{-1} x(\zeta). \tag{5}$$

The conventional frequency-domain interpretation emphasises the spectral characteristics of this inverse operation. In contrast, the present study focuses on its interpretation as a finite sequence of propagation operations acting on a finite-length seismic record.

Using the delay-operator representation introduced above, the inverse operator may formally be written as a Neumann series,

$$\left(I - rP(\zeta)\right)^{-1} = I + rP(\zeta) + r^2P^2(\zeta) + r^3P^3(\zeta) + \cdots.$$

At first sight, this expression appears to suffer from the same difficulties as the conventional inverse-filter formulation because it contains an infinite number of terms. However, the physical interpretation of $(P)$ as a propagation operator acting on a finite-length seismic record leads to a fundamentally different conclusion.

For a seismic record of finite duration $T$, repeated application of the propagation operator produces successive ghost propagations and reflections. Because the recorded wavefield occupies a finite temporal support, repeated application of the propagation operator eventually shifts all non-zero samples beyond the recording interval. When defined on this finite recording interval, the propagation operator is therefore nilpotent: there exists an integer (N) such that

$$P^{N+1}(\zeta) = 0,$$

and consequently,

$$P^{N+1}(\zeta)w(\zeta) = 0.$$

Under this condition, the Neumann series is not an infinite series whose convergence must be controlled; rather, it terminates after a finite number of terms. The resulting inverse operator is therefore a finite-duration operator and is represented by the finite expansion

$$\left(I - rP(\zeta)\right)^{-1} = I + rP(\zeta) + r^2P^2(\zeta) + \cdots + r^NP^N(\zeta). \tag{6}$$

The finite expansion is not introduced as an approximation but follows directly from the finite recording interval. As a consequence, the inverse operation is expressed as a finite sequence of propagation steps, providing the theoretical basis for the subsequent propagation-based wavefield decomposition framework.

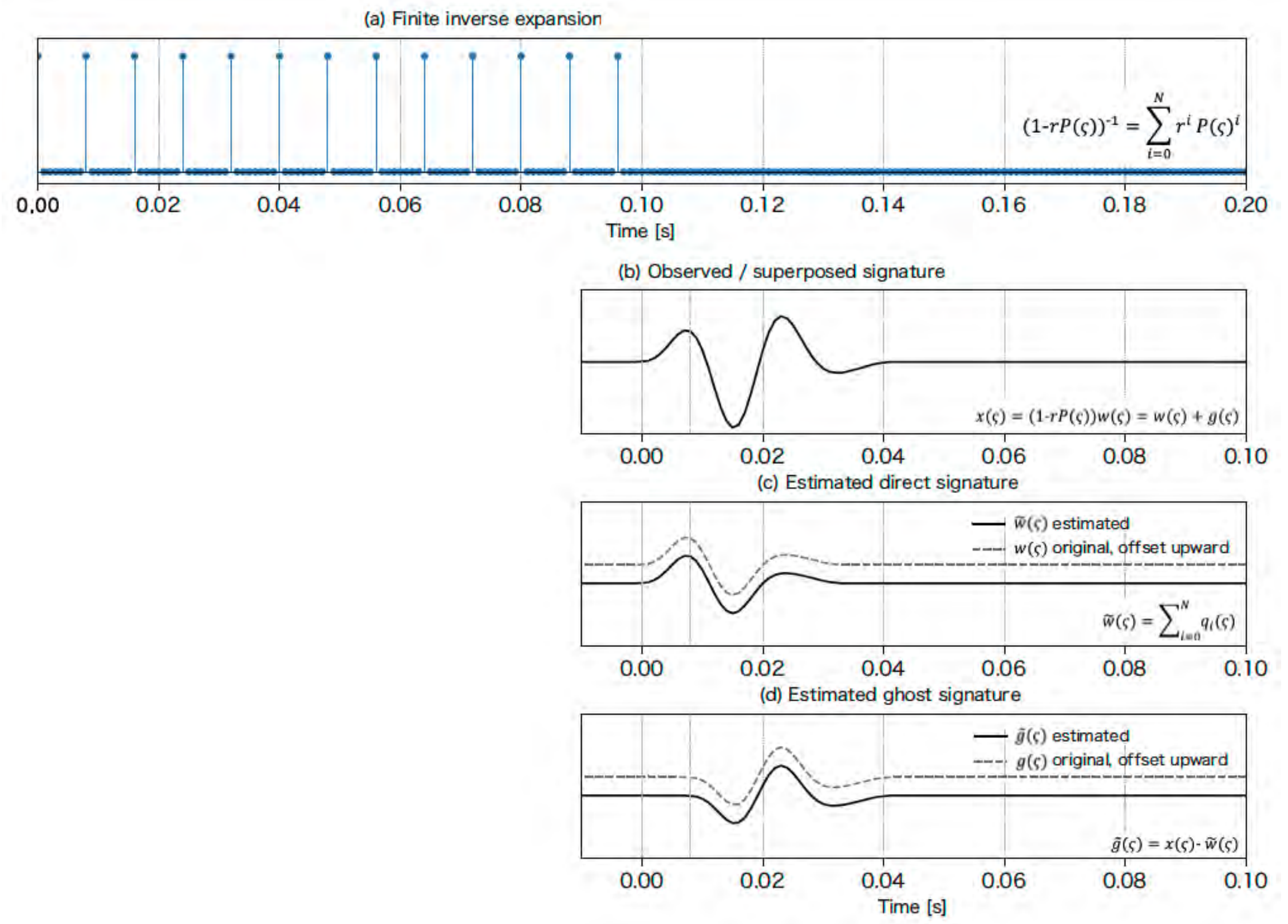

**Figure 2.** Demonstration of the finite inverse expansion used for decomposing the direct and ghost components from the observed signature. (a) Finite inverse operator ($h(t) = (I - E)^{-1} = I + E + E^2 + \cdots + E^N$), where ($E$) denotes the delay operator corresponding to the source ghost delay time ($\tau$). The operator is represented by a finite comb function because the finite recording interval renders the delay operator nilpotent beyond a finite order. Blue circles indicate zero-valued terms after termination of the finite expansion. (b) Observed (superposed) signature ($x(t) = (I - E) * w(t) = w(t) + g(t)$), where ($w(t)$) and ($g(t)$) denote the direct and ghost signatures, respectively. (c) Estimated direct signature ($\tilde{w}(t) = h(t) * x(t)$). (d) Estimated ghost signature ($\tilde{g}(t) = x(t) - \tilde{w}(t)$). Dashed curves in (c) and (d) represent the original direct and ghost signatures, respectively, shifted upward for visual comparison. The recovered signatures are identical to the originals within numerical precision, demonstrating that the finite inverse expansion decomposes the direct and ghost components for the synthetic example shown here. A 50-Hz minimum-phase Ricker wavelet is used for the source at a depth of 6 m, corresponding to a ghost delay of 8 ms relative to the direct arrival. For the 0.1-s record length used in this example, the maximum order of the finite expansion is $N = 12$, corresponding to 13 terms in total.

## 2.4 Recursive adaptive decomposition

The finite inverse expansion derived in the previous section may alternatively be interpreted as a recursive process rather than as a finite polynomial. Introducing a sequence of partial wavefields $q_i$, the direct signature is represented by

$$w(\zeta) = \sum_{i=0}^{N} q_i(\zeta),$$

where the successive components are generated recursively. Let the initial wavefield be

$$q_0(\zeta) = x(\zeta).$$

Introducing an auxiliary wavefield

$$u_i(\zeta) = \left(1 + rP(\zeta)\right) q_i(\zeta),$$

the recursive update is written as

$$q_{i+1}(\zeta) = u_i(\zeta) - q_i(\zeta).$$

Combining the above two relations immediately yields

$$q_{i+1}(\zeta) = rP\,(\zeta) q_i(\zeta), \tag{7}$$

which is precisely the recursive representation of the finite inverse expansion given by Equation (6). The finite Neumann expansion is therefore evaluated sequentially rather than simultaneously, with each recursive step corresponding to one additional propagation-and-reflection cycle. Thus, the ghost-free wavefield is obtained as the accumulated sum of the recursively generated partial wavefields. Although the synthetic example shown in Figure 2 uses a source signature for clarity, the recursive formulation is equally applicable to arbitrary seismic wavefields, including reflected events. The recursive formulation provides the theoretical basis for the recursive predictive decomposition developed in this study.

The recursive relation admits a straightforward physical interpretation. The initial component $q_0$ corresponds to the observed wavefield. The first recursive component $q_1$ represents the observed wavefield after one propagation to the free surface and one free surface reflection. Likewise, $q_2$ represents one additional propagation-and-reflection cycle applied to $q_1$. The same process continues successively until the finite expansion terminates naturally owing to the nilpotent property of the propagation operator. The auxiliary wavefield $u_i$ represents the wavefield after local ghost correction at the current recursive stage.

The recursive formulation described above assumes that the propagation operator $P$ is known a priori. In practice, however, the effective ghost response may be influenced by observational noise, source signature, local wavefield complexity, and small variations in ghost delay time. Consequently, the deterministic ghost operator $rP$ is replaced by an adaptively estimated prediction operator,

$$u_i(\zeta) = \big(1 + \alpha_i(\zeta)P_i(\zeta)\big)q_i(\zeta),$$

where $\alpha_i$ and $P_i$ denote the reflection-related operator and the corresponding propagation operator, respectively. When the effects of noise and other uncertainties are negligible, $\alpha_i$ reduces to the known free surface reflection coefficient $r = 1$ for the ideal pressure-release boundary considered here, and $P_i$ corresponds to the wavefield extrapolator for the known source or receiver depth $d$. When such effects cannot be neglected, $\alpha_i$ is represented by a polarity-reversed prediction-error operator and $P_i$ by the corresponding discrete prediction-delay operator associated with the known depth $d$. These operators may be estimated at $i = 0$ and retained throughout the recursive decomposition or, where appropriate, re-estimated at successive iterations from the autocorrelation of the evolving residual wavefield. In either case, their product provides a data-adaptive realisation of the deterministic ghost operator $rP$. The particular adaptive prediction scheme adopted in the present study is intended solely as one practical realisation of the deterministic framework; alternative ghost-decomposition algorithms may equally be incorporated within the same theoretical formulation.

The finite inverse expansion implies a comb-like propagation structure whose autocorrelation preserves the characteristic ghost delay. This observation explains why predictive decomposition naturally converges towards the ghost delay rather than requiring it to be prescribed a priori, thereby providing the theoretical basis for adaptive prediction.

The practical consequence of this recursive formulation is illustrated in Figure 2. Figure 2(a) shows the finite inverse operator represented by a comb function whose spacing corresponds to the ghost delay time ($\tau$). The propagation operator is conceptually identical to the discrete propagation operator introduced by Goupillaud (1961), in which wave propagation is represented as successive time-shift operations through layered media. Because the propagation operator is nilpotent when defined over the finite recording interval, the recursive expansion terminates naturally after a finite number of propagation steps. Applying the recursively constructed finite inverse operator to the noisy observed wavefield yields the direct-signature estimate

$$\tilde{w}(\zeta) \approx \sum_{i=0}^{N} q_i(\zeta),$$

shown in Figure 2(c). The ghost signature is subsequently obtained by subtraction,

$$\tilde{g}(\zeta) = x(\zeta) - \tilde{w}(\zeta),$$

as shown in Figure 2(d). The reconstructed direct and ghost signatures are indistinguishable from the original synthetic waveforms within numerical precision, demonstrating that the recursive formulation provides a realisation of the finite inverse expansion for the finite-length record considered here.

## 2.5 Practical recursive implementation

The adaptive propagation operator introduced in the previous section provides a practical realisation of the deterministic recursive formulation. Rather than prescribing a fixed propagation operator solely from the acquisition geometry, the recursive implementation may use either the deterministic operator defined by the known acquisition geometry or data-adaptive prediction operators estimated from the wavefield. These operators may be estimated once and then applied throughout the recursive decomposition or, where appropriate, re-estimated from the evolving residual wavefield during successive iterations. This flexibility allows the recursive implementation to accommodate observational noise, wavelet distortion, and local variations in effective ghost delay.

Rather than directly reconstructing the ghost-free wavefield, each recursive prediction step estimates only the local propagation-and-reflection process contained in the current residual wavefield. The accumulated sequence of these locally estimated operators progressively realises the finite inverse expansion derived from Equation (6). Consequently, the recursive predictive decomposition preserves the physical interpretation of the deterministic formulation while naturally adapting to local wavefield variations.

In practice, the local prediction operator $\alpha_i(\zeta)$ is obtained from a finite-length prediction error filter estimated from the current residual wavefield. Let the observed signature be represented by

$$q_0(\zeta) = x(\zeta).$$

A local prediction operator $\alpha_i(\zeta)$ is estimated from the current residual and is used to construct the partial estimate $u_i(\zeta)$, representing the wavefield component decomposed at the $i$-th recursive step,

$$u_i(\zeta) \quad = \quad (1 + \alpha_i(\zeta) P_i(\zeta)) \;\; q_i(\zeta).$$

The residual is then updated according to

$$q_{i+1}(\zeta) \quad = \quad u_i(\zeta) - q_i(\zeta),$$

which may also be written as

$$q_{i+1}(\zeta) \quad = \quad \alpha_i(\zeta) P_i(\zeta)\, q_i(\zeta). \tag{8}$$

The same procedure is subsequently applied to the updated residual. Repeated application of the local prediction operator therefore generates a sequence of residual components,

$$q_0(\zeta), \quad q_1(\zeta), \quad q_2(\zeta), \quad \ldots, \quad q_N(\zeta).$$

The final direct-signature estimate is obtained by summing all partial contributions,

$$\hat{w}(\zeta) = \sum_{i=0}^{N} q_i(\zeta). \tag{9}$$

Accordingly, the recursive formulation provides a practical realisation of the finite inverse expansion expressed by Equation (6). In the idealised case, the product $\alpha_i P_i$ reduces to the deterministic ghost operator $rP$, and the recursive decomposition becomes equivalent to the finite expansion

$$\left(I - rP(\zeta)\right)^{-1} = I + rP(\zeta) + r^2P^2(\zeta) + \cdots + r^NP^N(\zeta).$$

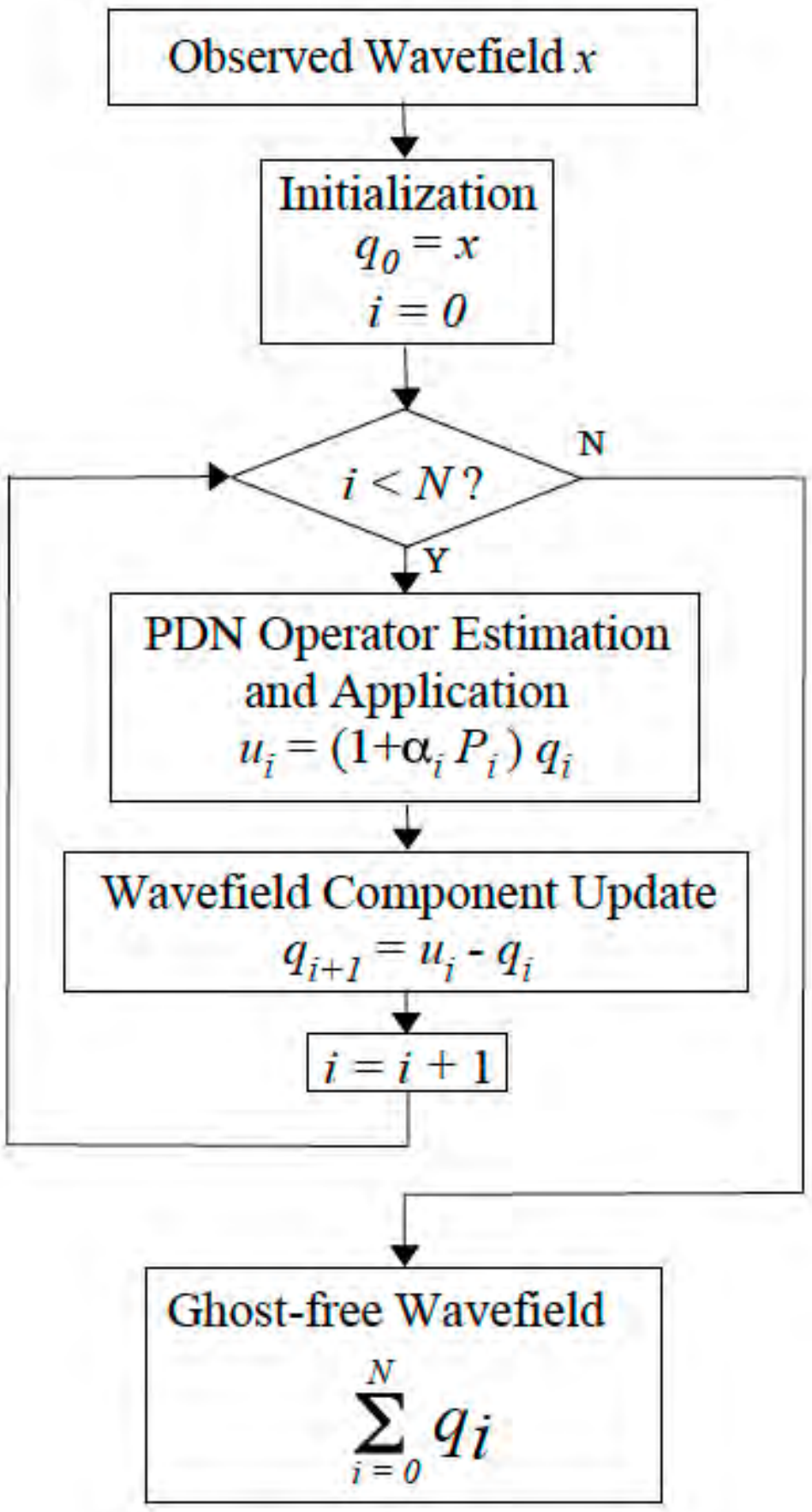

**Figure 3.** Flowchart of the recursive predictive decomposition algorithm derived from the recursive formulation of the finite inverse expansion. The observed wavefield is initialised as $q_0 = x$, after which the predictive deconvolution (PDN) operator shown in the figure is estimated and applied recursively to the current residual wavefield. The polarity-reversed prediction error operator $\alpha_i$ and the corresponding local delay operator $P_i$ may either be retained after the initial estimation or re-estimated during successive iterations to accommodate changes in the residual wavefield. The residual wavefield is then updated according to the recursive relation, and the procedure is repeated until the number of iterations $N$, naturally determined by the finite recording interval and the ghost-propagation delay, is reached. The final ghost-free wavefield is obtained by summing all recursively generated partial wavefields,
$\hat{w}(t) \approx \sum_{i=0}^{N} q_i(t)$.
The recursive implementation provides a practical, data-adaptive realisation of the finite inverse expansion derived in Section 2.3.

The practical implementation described above naturally leads to a comparison with predictive decomposition. Conventional predictive decomposition is commonly implemented as a single prediction step and may therefore be interpreted as retaining only the first recursion contribution of the finite inverse expansion. The recursive formulation extends this concept by repeatedly applying the prediction-and-subtraction process to the residual wavefield, thereby incorporating higher-order propagation terms. As a result, the recursive predictive decomposition, as illustrated in Figure 3, provides a practical approximation of the finite inverse expansion in the presence of noise, wavelet distortion, and modeling uncertainties.

The recursive implementation described above provides a practical means of decomposing the direct and ghost wavefields. Once these physically meaningful wavefield components have been obtained, they may be interpreted and manipulated individually. The decomposed direct and ghost wavefields provide the basis for the mirror-image interpretation and survey-sinking framework developed in the following sections.

**2.6 Mirror-image interpretation**

An alternative physical interpretation can be obtained in the spatial domain by introducing mirror-image sources and receivers with respect to the free surface, analogous to the mirror-image concepts previously employed in seismic imaging (e.g., Jamali Hondori et al., 2019). In this view, the ghost wavefields recorded below the surface can be regarded as originating from virtual sources and receivers located at mirrored positions above the free surface. This mirror-image representation provides a physically intuitive framework in which free surface ghosts are treated as additional wavefield components rather than as distortions of the primary signal. To use these components constructively, however, the direct and ghost components must first be decomposed from the observed signature.

The mirror-image interpretation provides more than a geometric description of free surface ghosts. Once the direct and ghost wavefields have been decomposed, the ghost components may be interpreted as wavefield components generated by virtual sources and receivers located at the mirrored positions. Accordingly, the decomposed ghost wavefields contain complementary information and are physically equivalent to additional wavefield components rather than redundant replicas of the direct wavefield. Because these virtual observations originate from mirrored source and receiver locations, they can be propagated to a common acquisition geometry through survey sinking before being superposed with the corresponding ghost-free wavefields. This physical interpretation establishes the conceptual foundation for the constructive ghost reuse strategy developed in the following sections.

### 2.7 Constructive reuse of ghost wavefields

Following the mirror-image interpretation introduced above, the decomposed ghost wavefields may be transformed into equivalent ghost-free wavefields through polarity reversal and propagation correction. In conventional deghosting workflows, the estimated ghost component is regarded as an unwanted by-product and is discarded after the direct signature has been recovered. In the present approach, however, the ghost estimate is treated as an additional wavefield component that may be constructively reused to improve the quality of the reconstructed direct signature.

Because the ghost component is represented by a delayed replica of the direct signature with opposite polarity, a second estimate of the direct signature can be obtained by reversing the polarity of the estimated ghost and compensating for the ghost delay time. The resulting ghost-derived estimate contains the same physical signal as the direct estimate, but generally exhibits different error characteristics because it is recovered through a different prediction pathway.

The final direct-signature estimate is obtained by wavefield superposition of the direct estimate and the corrected ghost estimate,

$$\hat{w}_s(\zeta) = \hat{w}_{\mathrm{d}}(\zeta) + \hat{w}_g(\zeta), \tag{10}$$

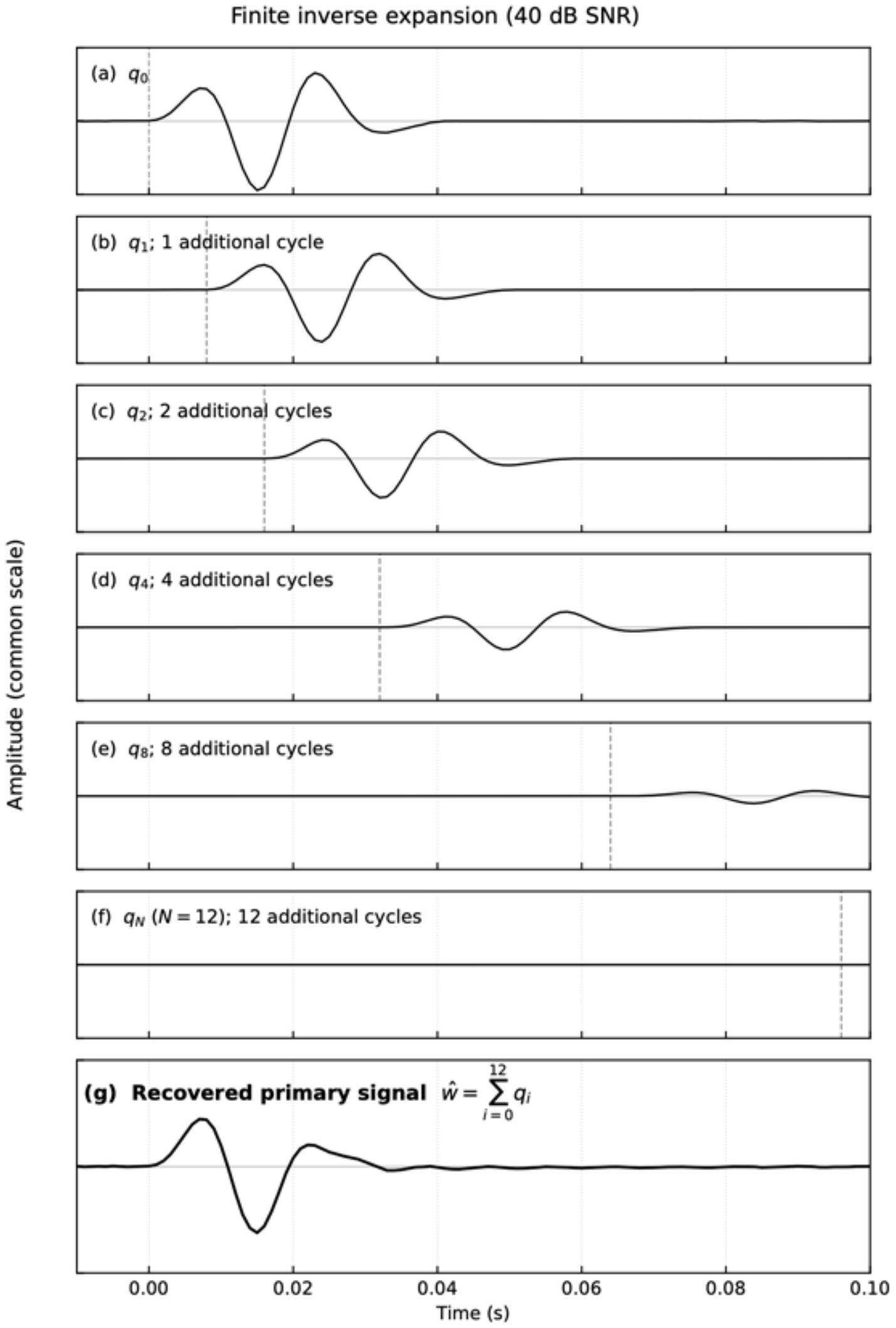

**Figure 4.** Illustration of the recursive predictive decomposition for a synthetic trace with 40 dB signal-to-noise ratio. Panels (a)–(f) show representative partial wavefields $q_i$ obtained from the finite inverse expansion: $q_0$, $q_1$, $q_2$, $q_4$, $q_8$, and $q_N$ $(N = 12)$, respectively. Each $q_i$ is the $i$-th term of the expansion and represents a partial wavefield that has undergone one additional free surface propagation-and-reflection cycle relative to $q_{i-1}$. The dashed vertical line marks the corresponding cumulative propagation delay, demonstrating how the ghost contribution is progressively displaced toward, and ultimately beyond, the end of the finite recording interval. Panel (g) shows the recovered primary wavefield, $\hat{w} = \sum_{i=0}^{N} q_i$, obtained by coherently summing all partial wavefields. All panels use identical time and amplitude scales.

where $(\hat{w}_{\mathrm{d}}(\zeta))$ denotes the direct-signature estimate obtained from the recursive predictive decomposition and $(\hat{w}_g(\zeta))$ denotes the polarity-reversed and time-shifted ghost estimate. No amplitude scaling was applied to the ghost-derived estimate prior to stacking. The theoretical formulation considers only wavefield superposition and therefore does not include amplitude normalisation. Any normalisation applied to the final estimate $\hat{w}_s(\zeta)$ is implementation-dependent and does not affect the theoretical formulation. Equation (10) describes the one-dimensional source-side case. When both source- and receiver-side ghosts are considered, the decomposition is applied sequentially to the two sides, leading to the four real/mirrored source–receiver combinations introduced in the multidimensional formulation below.

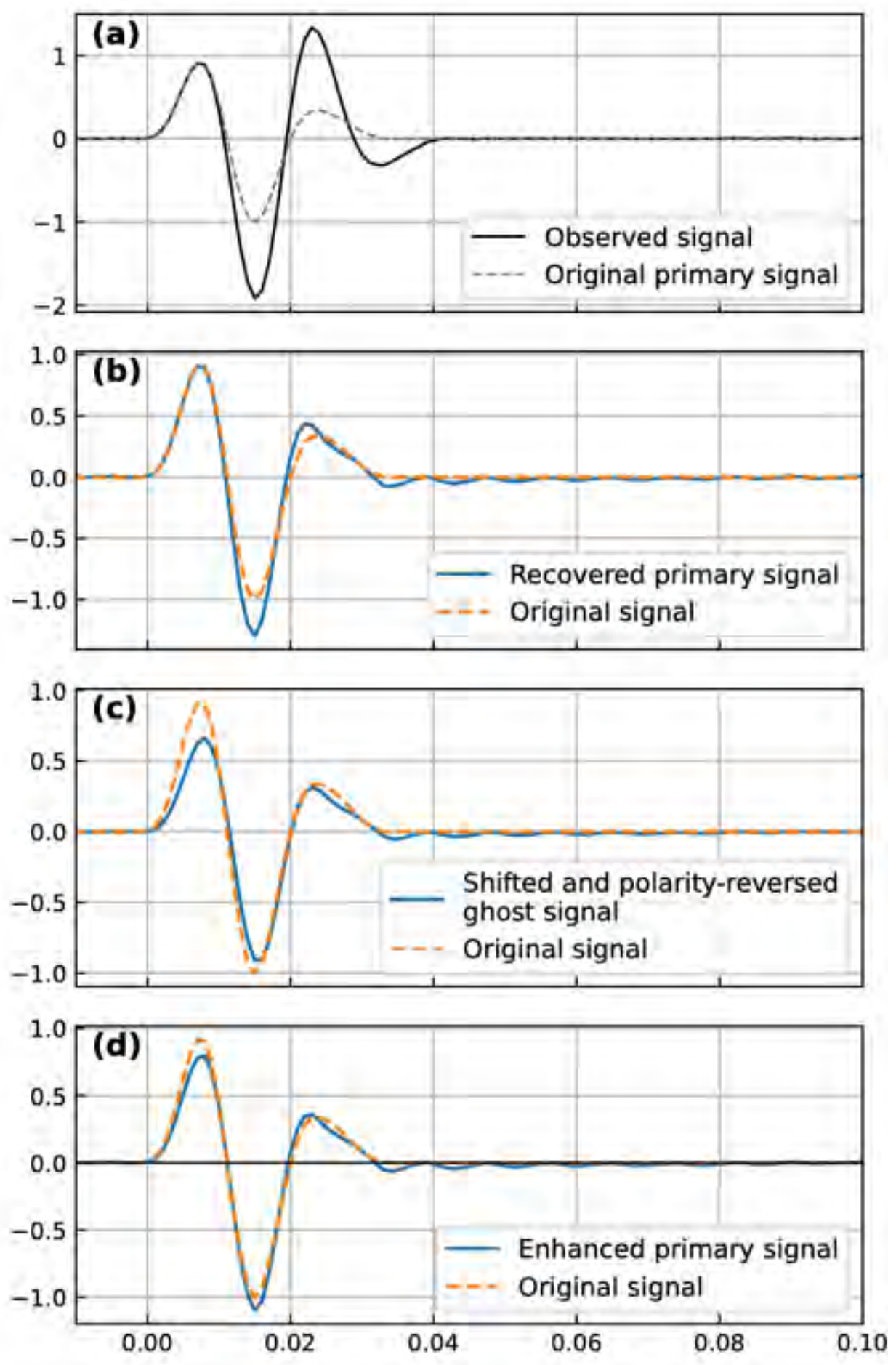

**Figure 5.** Example of direct-signature reconstruction and constructive reuse of ghost signature for an input S/N ratio of 20 dB. (a) Observed superposed signature contaminated by additive random noise. (b) Estimated direct signature obtained using the recursive predictive decomposition. (c) Estimated ghost signature after polarity reversal and correction for the ghost delay for comparison with the direct signature. (d) Averaged estimate obtained by averaging the estimated direct signature and the corrected ghost estimate. No amplitude scaling was applied to the ghost estimate prior to averaging. Despite the absence of amplitude balancing, the averaged estimate exhibits a closer agreement with the original direct signature than either estimate individually.

Figure 4 illustrates how the partial wavefield $q_i$ evolves at each step of the recursive decomposition. Each successive component undergoes one additional free surface propagation-and-reflection cycle, resulting in a progressive delay by the ghost delay time and, in this example, a gradual decrease in amplitude. Figure 5 presents an example with an input S/N ratio of 20 dB. The direct estimate and the ghost-derived estimate both reproduce the principal features of the original signature. For quantitative comparison with the original wavelet, their average provides a closer approximation to the original direct signature than either estimate individually.

**Table 1.** Comparison of the RMS error and correlation coefficient of the estimated direct signature and the stacked direct-ghost estimate for different input S/N ratios. The stacked estimate consistently exhibits a lower RMS error and a higher correlation with the original direct signature than the direct estimate alone. The observed improvement is approximately 3 dB at low input S/N levels and exceeds 4 dB at high input S/N levels, indicating that the errors in the direct and ghost estimates are partially complementary.

| Input S/N (dB) | Direct RMS Error | Stacked RMS Error | Improvement (dB) | Direct Corr. | Stacked Corr. |
|---|---|---|---|---|---|
| 10 | 0.1375 | 0.0969 | 3.04 | 0.8419 | 0.8738 |
| 20 | 0.0611 | 0.0425 | 3.15 | 0.9655 | 0.9731 |
| 30 | 0.0460 | 0.0297 | 3.81 | 0.9834 | 0.9876 |
| 50 | 0.0442 | 0.0276 | 4.08 | 0.9859 | 0.9899 |
| 100 | 0.0442 | 0.0276 | 4.10 | 0.9859 | 0.9900 |

The quantitative results are summarised in Table 1. The stacked estimate consistently exhibits a lower RMS error and a higher correlation coefficient than the direct estimate alone over the entire range of input S/N ratios examined. The observed improvement is approximately 3 dB at low input S/N levels and exceeds 4 dB at high input S/N levels. This behaviour suggests that the residual errors contained in the direct and ghost estimates are partially complementary rather than purely random, allowing the stacking process to provide a greater reduction in error than expected from simple averaging of two statistically independent estimates.

## 2.8 Survey sinking to a common acquisition geometry

Once the direct and ghost wavefields have been decomposed, each component may be propagated to a common acquisition geometry corresponding to the real source and receiver positions. This survey sinking operation restores the kinematic consistency among the decomposed wavefields, allowing them to be combined constructively through wavefield superposition. As a result, this approach can be implemented as a preprocessing operation that generates additional physically consistent wavefield components without modifying the original acquisition geometry.

Under the mirror-image interpretation, the decomposed ghost wavefields may be regarded as observations associated with corresponding virtual source and receiver locations. Combined with reciprocity (Wapenaar, 1998), this propagation enables the construction of equivalent observations that would have been recorded by a survey physically located at the mirrored positions. In this sense, the proposed propagation-based processing framework may be interpreted as a virtual survey sinking operation achieved entirely through wavefield processing.

Unlike conventional deghosting, which aims solely at recovering the primary wavefield, survey sinking enables both direct and ghost wavefield components to be constructively reused. As a result, available wavefield components can be brought to a common acquisition geometry and superposed without requiring additional acquisition, thereby enhancing coherent signal energy while preserving the physical consistency of the recorded wavefield. In the present study, survey sinking is used not for imaging itself but for restoring a common acquisition geometry prior to wavefield superposition. The proposed framework is naturally extended to multidimensional wavefields through wavefield extrapolation (Gazdag, 1978; Berkhout, 1980) and reciprocity (Wapenaar, 1998).

## 3. Multidimensional extension and physical interpretation

### 3.1 Multidimensional formulation

The finite inverse expansion developed in Section 2 can be naturally extended to multidimensional wavefields by generalising the temporal delay representation of the propagation operator to a multidimensional wavefield-extrapolation operator. Whereas the one-dimensional formulation was expressed using the z-transform variable $\zeta$, the multidimensional formulation is represented in the frequency–wavenumber domain $(k_x, k_y, \omega)$. In multidimensional media, propagation is no longer described solely by a time delay. Instead, the corresponding propagation operator is expressed as a wavefield-extrapolation operator. For propagation over a vertical distance $d$, the phase-shift extrapolation operator $P$, represented by a discrete delay in the preceding one-dimensional formulation, may be written as

$$P(k_x, k_y, \omega; d) = \exp(2ik_z d), \tag{11}$$

where $k_z$ denotes the vertical wavenumber,

$$k_z = \sqrt{\left(\frac{\omega}{V}\right)^2 - k_x^2 - k_y^2}, \tag{12}$$

where $\omega$ is the angular frequency, $V$ is the acoustic velocity of seawater, $k_x$ and $k_y$ are the horizontal wavenumbers. This expression corresponds to the phase-shift representation of wavefield extrapolation (Gazdag, 1978) and provides a multidimensional counterpart of the one-dimensional delay operator. The one-dimensional formulation is recovered for normal incidence ($k_x = k_y = 0$), for which $k_z = \omega/V$, so that

$$P = \exp(2i\omega d/V) = \exp(i\omega\tau),$$

with $\tau = 2d/V$. Thus, the familiar one-dimensional phase-delay representation is recovered as the normal-incidence limit of the multidimensional propagation operator.

More generally, this replacement of the one-dimensional discrete delay operator by a vertical-wavenumber extrapolation operator connects the finite inverse expansion developed in Section 2 to multidimensional wavefield extrapolation theory (Berkhout, 1980).

The ghost operator may therefore be expressed as the product of a propagation operator and a free surface reflection operator,

$$G = RP,$$

where $R$ represents the free surface reflection operator generalised from the scalar reflection coefficient magnitude r in the one-dimensional case, with the polarity reversal represented separately by the sign convention adopted in Section 2. Accordingly, the finite inverse expansion may be interpreted as a finite sequence of wavefield propagations and free surface reflections rather than a simple series of time delays.

Replacing the one-dimensional delay operator in Equation (6) by its multidimensional propagation counterpart immediately yields the corresponding finite inverse expansion,

$$(I - RP)^{-1} = I + RP + (RP)^2 + \cdots + (RP)^N \tag{13}$$

Equation (13) therefore represents the multidimensional counterpart of the finite inverse expansion developed in Section 2. Although the mathematical form remains unchanged, the propagation operator $P$ represents wavefield extrapolation between the source or receiver location and the free surface in multidimensional media. The individual terms of the expansion therefore acquire a direct physical interpretation as distinct propagation paths.

**3.2 Physical interpretation**

The first term, $I$, represents the direct wavefield. The second term, $RP$, corresponds to a single propagation-and-reflection cycle between the source or receiver location and the free surface. Higher-order terms, $(RP)^2$, $(RP)^3$, ..., represent successive propagation-and-reflection cycles. The multidimensional finite inverse expansion may therefore be interpreted as a finite summation of physically meaningful propagation paths.

This interpretation differs from the conventional frequency-domain view of ghost removal. Instead of constructing an inverse filter whose properties are determined by the locations of poles and zeros, the decomposition is expressed as a finite sequence of wavefield propagations and reflections. The resulting formulation therefore provides a direct physical interpretation of multidimensional wavefield propagation rather than merely a mathematical inverse-filtering operation.

### 3.3 Reciprocity

The explicit representation of the propagation operator enables reciprocity to be incorporated directly into the formulation (Wapenaar, 1998). Source- and receiver-side ghosts are therefore interpreted as reciprocal manifestations of the same propagation process, differing only in observational perspective. This interpretation provides the physical basis for treating both within a common propagation-based framework. The multidimensional formulation therefore provides not only a mathematical extension of the finite inverse expansion, but also a physically interpretable framework in which decomposed ghost wavefields can be constructively reused for coherent signal enhancement. The physical interpretation of Equation (13) is illustrated by the synthetic example shown in Figure 6.

Figure 6 demonstrates that the finite inverse expansion decomposes the observed wavefield into an estimated primary wavefield and an estimated ghost wavefield. The ghost-reused wavefield is subsequently constructed according to Equation (10) by superposing the estimated primary wavefield with the corrected ghost-derived estimates. The corresponding amplitude spectra are shown in Figure 7. The ghost-induced spectral notches are substantially recovered, while the ghost-reused wavefield exhibits an approximately 6 dB increase in coherent reflection amplitude relative to the true primary over most of the usable frequency band. The quantitative performance of the proposed ghost-reuse procedure is summarised in Table 2. Relative RMS error, waveform correlation, and the measured ghost-reuse signal gain are presented for input S/N ratios of 20, 30, and 40 dB. As the input S/N increases, the measured ghost-reuse signal gain converges toward the theoretical value of approximately 6 dB, while both the relative RMS error and waveform correlation exhibit consistent improvement.

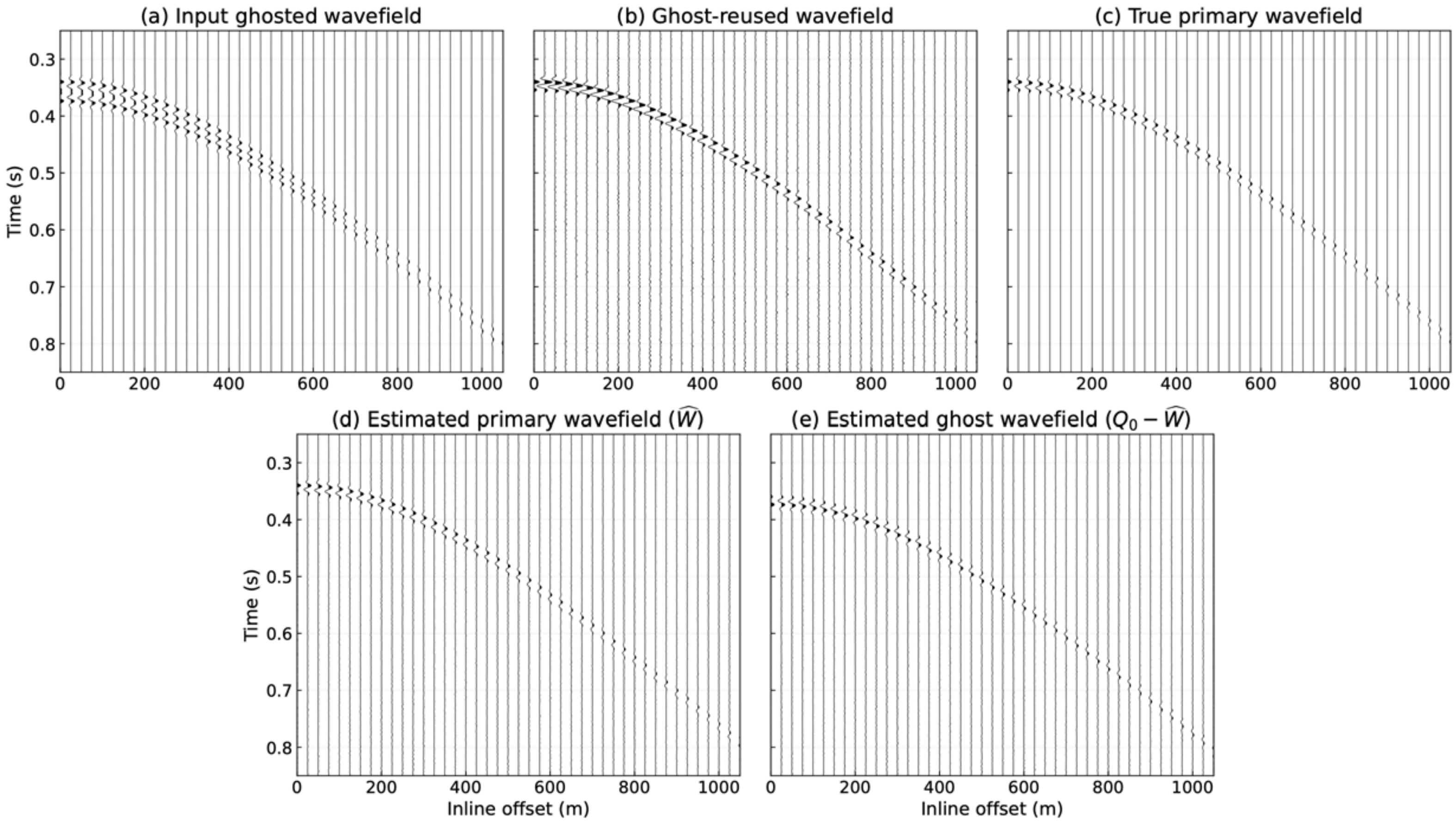

**Figure 6.** Synthetic 2-D wavefields illustrating ghost decomposition and reuse based on Equation (10). (a) Input ghosted wavefield ($Q_0$), (b) ghost-reused wavefield, (c) true primary wavefield, (d) estimated primary wavefield ($\widehat{W}$), and (e) estimated ghost wavefield ($Q_0 - \widehat{W}$). The ghost-reused wavefield was obtained by superposing the estimated primary wavefield with the ghost-derived estimate after polarity reversal and propagation correction, according to Equation (10). All panels are displayed using the same amplitude scale for direct visual comparison.

**Table 2.** Ghost-reuse Signal Gain was measured within the primary-hyperbola signal window relative to the true primary wavefield. As the input S/N increases, the measured gain converges toward the theoretical value of approximately 6 dB expected for coherent ghost reuse. The larger gain observed at 20 dB S/N reflects the contribution of residual noise and processing-related energy within the measurement window.

| Input S/N (dB) | Primary Relative RMS Error | Ghost-reused Relative RMS Error | Improvement (dB) | Primary Correlation | Ghost-reused Correlation | Measured Ghost-reuse RMS Gain (dB) |
|---|---|---|---|---|---|---|
| 20 | 0.578 | 0.568 | 0.145 | 0.866 | 0.870 | 10.5 |
| 30 | 0.187 | 0.183 | 0.189 | 0.983 | 0.984 | 6.8 |
| 40 | 0.072 | 0.068 | 0.493 | 0.997 | 0.998 | 6.1 |

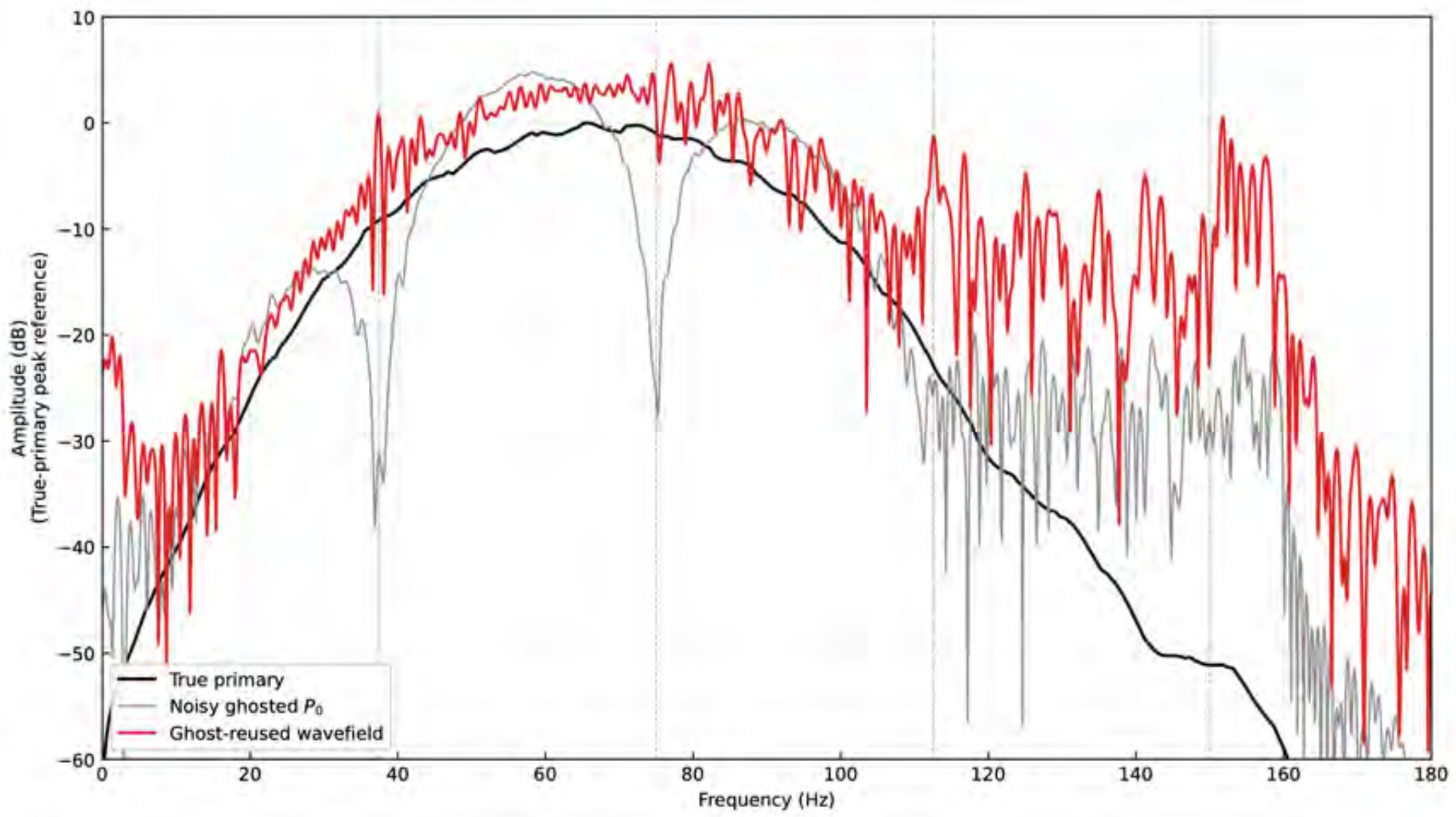

**Figure 7.** Amplitude spectra of the true primary, the input ghosted wavefield, and the ghost-reused wavefield corresponding to Figure 6. The ghost-reused wavefield partially restores spectral amplitudes at ghost-related notches introduced by the free surface ghost, exhibiting an approximately 6 dB increase in coherent reflection amplitude relative to the true primary over most of the displayed frequency range. All spectra were calculated using a 5% cosine taper.

## 4. Demonstration of the proposed propagation-based processing framework

### 4.1 Decomposition of direct and ghost signatures

The one-dimensional formulation developed in Section 2 provides a convenient framework for illustrating the finite inverse expansion and its recursive implementation. Source- and receiver-side ghosts are represented by delay operators of the same mathematical form, with their respective delay times determined independently by the source and receiver depths.

In multidimensional seismic data, however, the situation differs fundamentally. Source and receiver ghosts exhibit different spatial behaviours and are naturally observed in common-shot and common-receiver gathers, respectively. This additional spatial information allows the two ghost contributions to be treated separately while remaining connected through reciprocity (Wapenaar, 1998). The multidimensional dataset therefore provides a practical framework for the separate decomposition and constructive reuse of source- and receiver-side ghost wavefields.

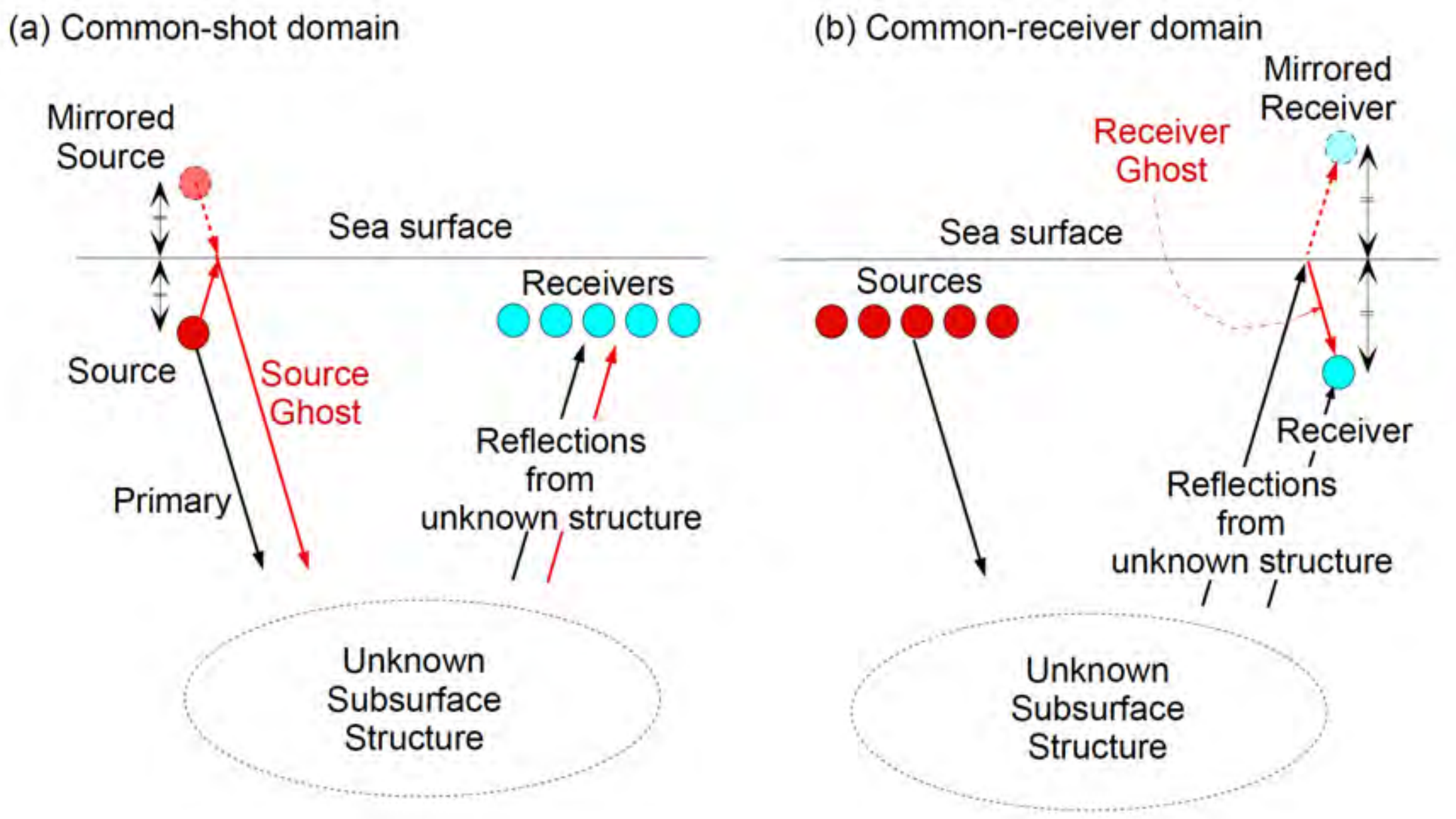

**Figure 8.** Conceptual illustration of source- and receiver-side ghost generation in multidimensional marine seismic acquisition. (a) Source ghosts represented by a mirrored source above the free surface in a common-shot gather. (b) Receiver ghosts represented by a mirrored receiver above the free surface in a common-receiver gather. Because ghost propagation is confined to the water column, the ghost propagation operator depends primarily on acquisition geometry and water sound velocity and does not require a model of the underlying geological structure.

An additional advantage of the proposed formulation is that ghost generation is confined to the water column between the acquisition system and the free surface. Consequently, the propagation operators depend primarily on the acquisition geometry and the acoustic velocity of seawater without requiring a subsurface velocity model.

This simplification remains valid throughout the recursive decomposition process. Even for higher-order terms of the finite inverse expansion, the associated propagation paths are restricted to repeated traverses between the free surface and the source or receiver locations. Neither the individual propagation operators nor their iterative accumulation requires a geological velocity model beneath the seafloor.

Prior to the multidimensional decomposition, appropriate source- and receiver-static corrections are assumed to have been applied so that the observed event moveout is consistent with the physical propagation behaviour considered in the decomposition. The multidimensional decomposition is performed in two successive stages. Source-side decomposition is first carried out in the common-shot domain, followed by receiver-side decomposition in the common-receiver domain (Figure 8). First, source ghosts are estimated and removed in the common-shot domain. Because source ghosts originate from propagation between the source location and the free surface, they appear coherently within each shot gather and can be isolated through the proposed recursive decomposition framework. The spatial-sampling requirements may differ between these two domains because the spatial sampling of common-shot gathers is governed by the receiver distribution, whereas that of common-receiver gathers is governed by the source distribution. The accuracy of the multidimensional decomposition therefore depends on the spatial sampling available in the corresponding processing domain. Spatial components that are insufficiently sampled may be affected by spatial aliasing and cannot be recovered simply by using a finer processing grid or CMP binning.

Spatial regularisation or interpolation may improve the representation of the recorded wavefield where appropriate, but cannot recover spatial-wavenumber information that was not adequately sampled in the original acquisition.

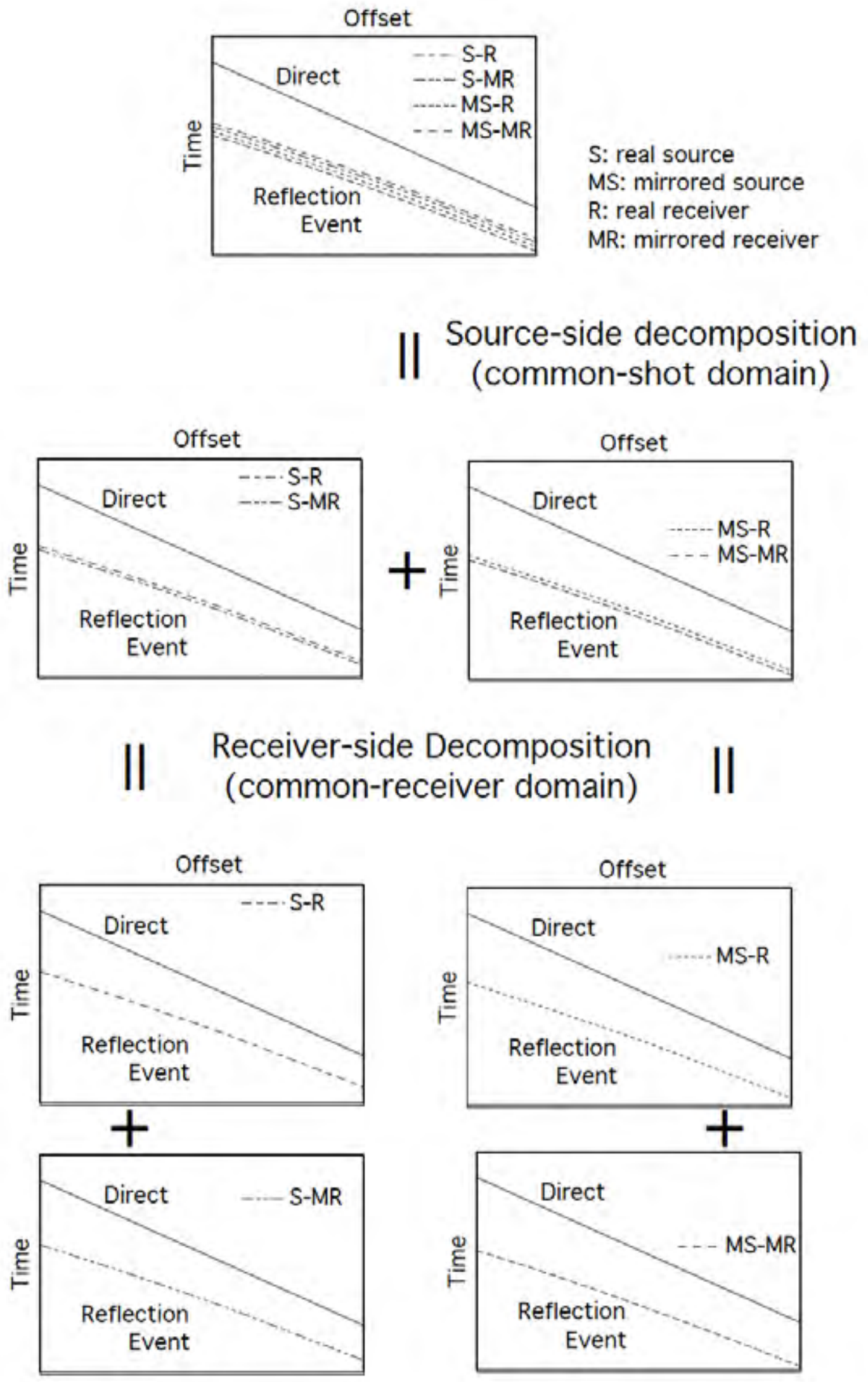

**Figure 9.** Conceptual workflow for multidimensional wavefield decomposition. The recorded wavefield is first decomposed in the common-shot domain into a real-source gather and a source-ghost gather. Each component is subsequently decomposed in the common-receiver domain, yielding four physically interpretable wavefield classes: S–R (real source–real receiver), S–MR (real source–mirrored receiver), MS–R (mirrored source–real receiver), and MS–MR (mirrored source–mirrored receiver). These decomposed wavefields form the basis for the constructive reuse framework developed in this study. The workflow shown here focuses on the decomposition and constructive reuse of reflected wavefields; the direct arrival and its closely overlapping source ghost are not treated separately in this conceptual illustration.

The resulting source-deghosted data are subsequently reorganised into common-receiver gathers. Receiver ghosts are then estimated and removed using the same recursive decomposition procedure. Reciprocity ensures that the underlying propagation process is equivalent, allowing the same formulation to be applied in both domains without modification.

Following source- and receiver-side decomposition, the recorded wavefield is decomposed into physically interpretable components. Source-side decomposition in the common-shot domain yields wavefields associated with the real and the mirrored sources. Subsequent receiver-side decomposition of each component produces four wavefield classes: (1) direct wavefields free of both source and receiver ghosts, (2) wavefields associated with the real source and mirrored receivers, (3) wavefields associated with the mirrored source and real receivers, and (4) wavefields originating from the mirrored source at the mirrored receivers. These components form the basis for the survey sinking and wavefield-superposition procedures described in the following section.

## 4.2 Multidimensional wavefield decomposition and restoration

Figure 9 illustrates the conceptual decomposition of the recorded wavefield into four physically interpretable classes corresponding to combinations of real and mirrored sources and receivers. Source-side decomposition separates the recorded wavefield into components associated with the real and mirrored sources. Subsequent receiver-side decomposition further separates each component according to the real and mirrored receiver positions, yielding four wavefield classes: S-R, S-MR, MS-R, and MS-MR.

This decomposition enables the decomposed ghost wavefields to be treated as physically meaningful wavefield components rather than as unwanted energy to be discarded. Because each decomposed component is associated with a distinct acquisition geometry, the individual wavefields cannot be combined directly.

Figure 10 illustrates the corresponding restoration process. The decomposed wavefields are first restored to a common acquisition geometry through survey sinking. Once the geometrical consistency has been re-established, the wavefields are constructively combined by wavefield superposition to produce a reconstructed S-R gather. The decomposition and restoration procedures therefore form a unified framework in which decomposed ghost wavefields are transformed into physically consistent additional wavefield components rather than being discarded.

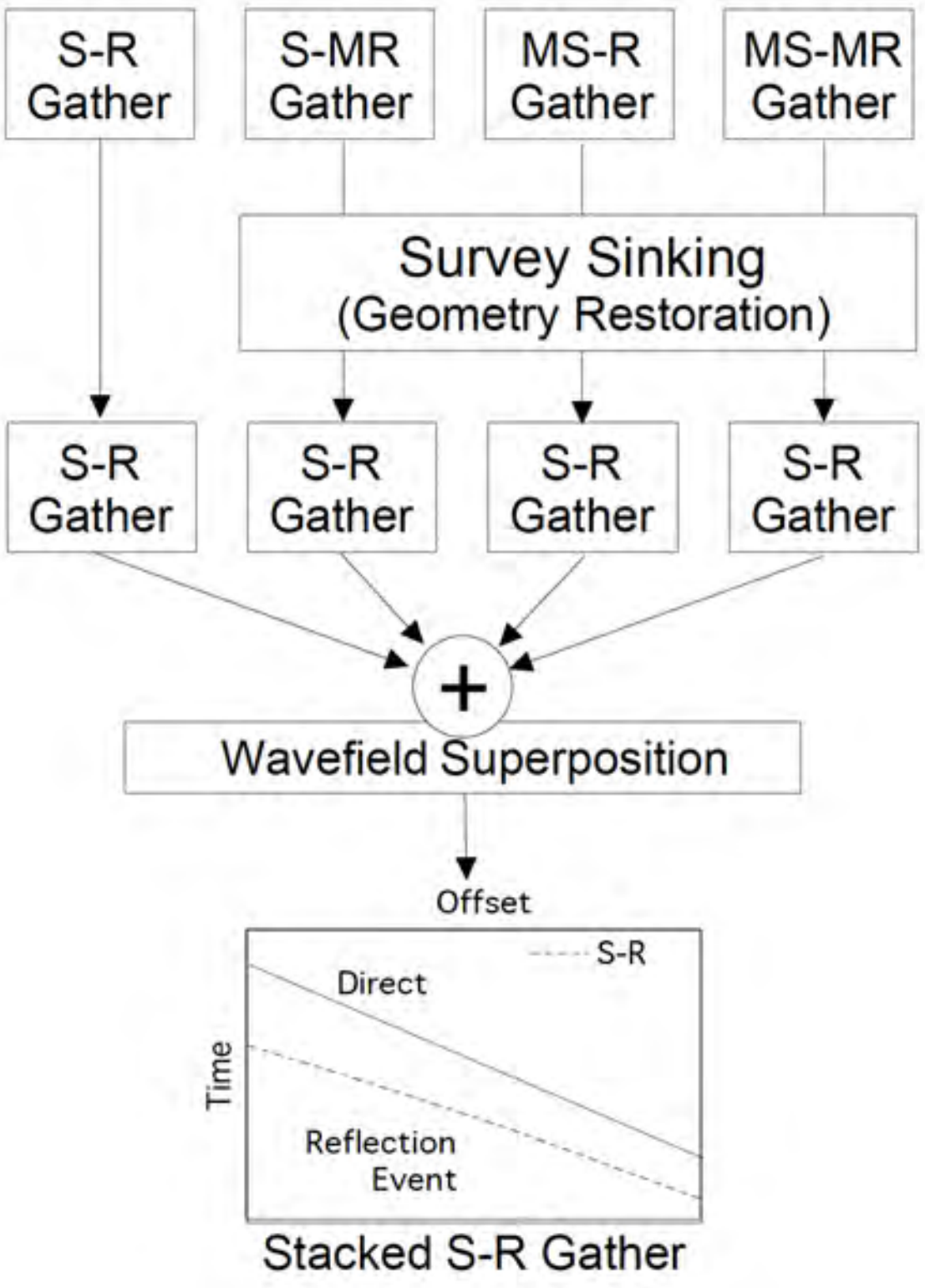

**Figure 10.** Conceptual workflow for multidimensional wavefield restoration through survey sinking and wavefield superposition. The four decomposed wavefields obtained from source- and receiver-side decomposition are transformed into a common real source–real receiver geometry through survey sinking (geometry restoration). Because all wavefields are restored to the same acquisition geometry, they can subsequently be combined by wavefield superposition to generate a stacked common-shot gather with enhanced signal quality.

### 4.3 Field-data demonstration

The field dataset analysed in this study was acquired during the marine seismic survey reported by Ozasa et al. (2025). The basic processing sequence is therefore not repeated here. Whereas Ozasa and Mikada (2025) primarily focused on the practical implementation of the proposed processing sequence, the present study provides a unified propagation-based interpretation that places the same processing sequence within a deterministic wave-propagation framework. This dataset provides a suitable test case because the relatively shallow source and streamer depths generate pronounced source and receiver ghosts.

The processing sequence followed the multidimensional decomposition and restoration framework described in the preceding sections. Source and receiver ghost wavefields were decomposed using multichannel predictive deconvolution following the recursive procedure described in Section 2.5. At each recursive step, an auxiliary wavefield component $u_i(\zeta)$ was estimated from the current residual $q_i(\zeta)$, and the updated residual wavefield $q_{i+1}$ was then used as the input to the subsequent step. This procedure was repeated until the propagation residual moved outside the finite recording interval. The resulting decomposed wavefield components were restored to a common acquisition geometry by survey sinking and combined through wavefield superposition. The present formulation provides a physical interpretation of this practical recursive procedure in terms of the finite inverse expansion and the nilpotent character of the propagation operator over a finite recording interval.

Following validation using the synthetic examples presented in Section 3, the same multidimensional processing framework was applied to field seismic data. The resulting field-data example is presented in Figure 11. Figures 11(a) and 11(b) compare the migrated seismic sections before and after application of the proposed framework, demonstrating enhanced coherent reflection amplitudes and improved reflector continuity. Representative seismic traces are shown in Figure 11(c), illustrating partial recovery of the ghost-distorted wavelet. The corresponding amplitude spectra are presented in Figure 11(d), showing partial filling of the ghost-related spectral notches together with improved low-frequency content. Overall, the field-data results demonstrate that the decomposed ghost wavefields can be constructively reused as physically meaningful additional wavefield components to improve image quality and extend the usable bandwidth without modifying the original acquisition geometry.

## 5. Discussion

### 5.1 Propagation-based interpretation of ghost wavefields

The present propagation-based framework builds upon recent advances in ghost decomposition and extends their role beyond conventional ghost attenuation. Rather

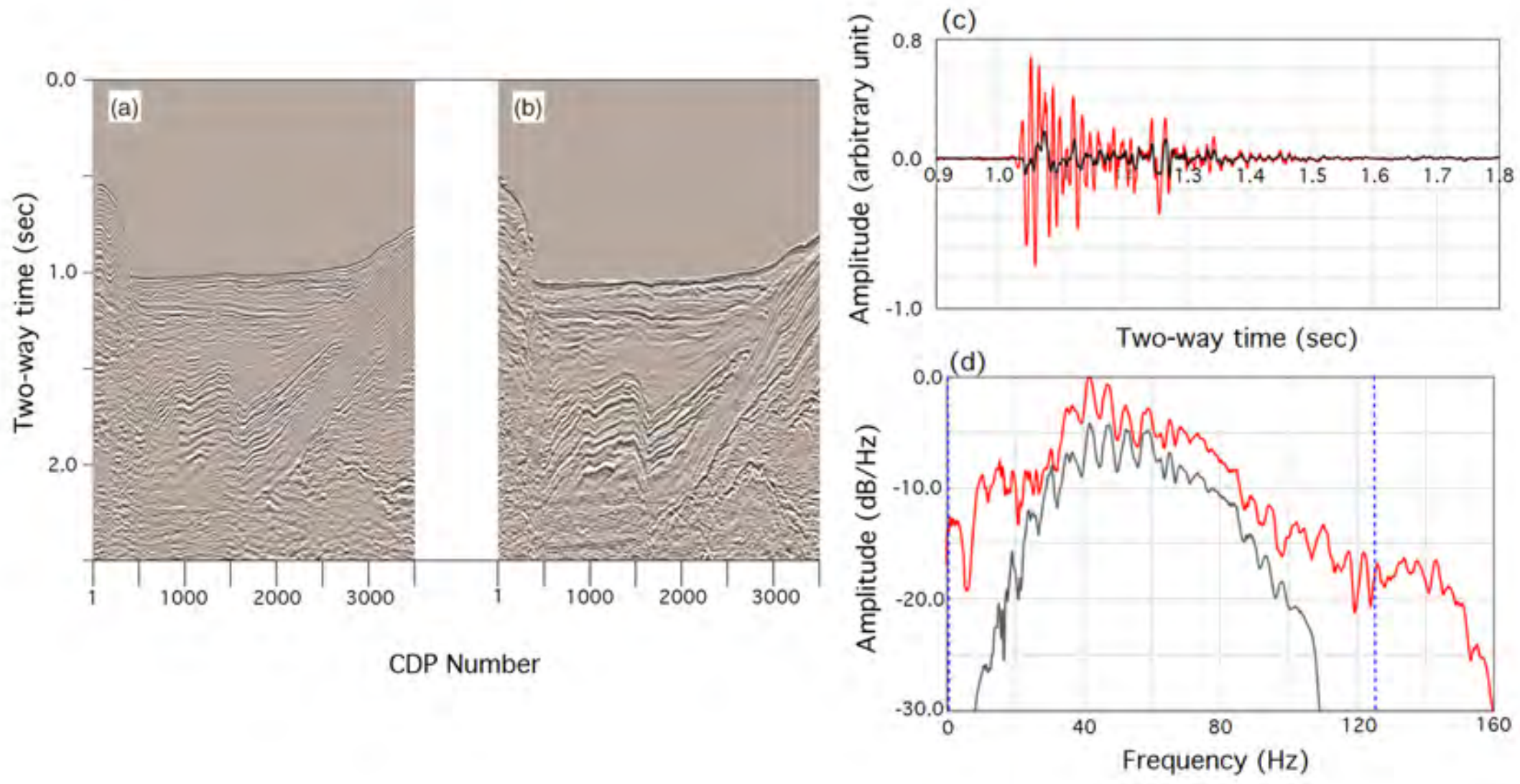

**Figure 11.** Field-data example demonstrating processing-driven signal enhancement through constructive reuse of free surface ghost energy, reproduced and modified from Ozasa and Mikada (2025).

Panels (a) and (b) show stacked CMP sections before and after processing, respectively, acquired using a shallow-towed airgun source and streamer both deployed at a depth of approximately 6 m, with a water sound velocity of 1.5 km/s. The streamer consisted of 168 hydrophone channels with an inline receiver spacing of 12.5 m, and the shot spacing was also 12.5 m, resulting in a nominal CMP spacing of 6.25 m. Source- and receiver-side free surface ghost components were decomposed using multichannel predictive deconvolution following the workflow illustrated in Figures 8–10. Panel (c) compares representative traces before (black) and after (red) ghost reuse and panel (d) shows partial restoration of both low- and high-frequency components attenuated by the free surface ghost, including spectral amplitudes at the ghost notches at 0 and 125 Hz. These components are recovered through constructive reuse of the decomposed ghost wavefields.

The original inline receiver and source spacings were both 12.5 m, giving a nominal CMP spacing of 6.25 m. Under the assumption of a two-dimensional subsurface structure, natural receiver-position variations caused by streamer feathering were exploited during CMP binning, and the final CMP sections were constructed with a spatial sampling interval of 3.125 m, allowing the subsequent migration to accommodate frequencies well above the 125-Hz ghost notch. This finer CMP sampling enables the higher-frequency components recovered through ghost reuse to be retained in the subsequent imaging process.

than focusing on surface-related multiples generated by repeated interactions between the free surface and the subsurface, the proposed framework interprets the free surface ghost as a deterministic propagation process occurring between the acquisition system and the free surface. From this perspective, the ghost wavefield is treated not as unwanted energy to be suppressed, but as a physically meaningful component that can be decomposed, interpreted, and constructively reused.

The underlying philosophy also differs from conventional suppression-based approaches in that the ghost wavefield is regarded as physically meaningful wavefield information rather than unwanted energy. Previous propagation-based processing and processing-based deghosting studies have demonstrated the effectiveness of wavefield extrapolation and recursive deghosting strategies (Amundsen et al., 2013; Ferber and Beasley, 2014). The present study differs in that the propagation-based framework is derived directly from the finite inverse expansion of the ghost operator and subsequently extended through mirror-image interpretation, survey sinking, and wavefield superposition. The proposed framework therefore provides a unified physical interpretation of ghost decomposition, restoration, and constructive reuse. In this respect, the present study shares a conceptual similarity with methodologies that seek to exploit multiply scattered wavefields rather than remove them, although the physical phenomena and practical objectives are fundamentally different. More broadly, the present formulation represents a shift from ghost suppression to ghost interpretation and constructive reuse. The source-side formulation also suggests a potential application to marine source-wavelet estimation, because the direct source signature can be estimated from the observed wavefield by exploiting the deterministic propagation relationship between the direct and free surface ghost components.

### 5.2 Practical implications for marine seismic data acquisition and processing

The propagation-based framework presented in this study suggests that the conventional strategy may deserve reconsideration. If free surface ghosts can be decomposed and constructively reused during preprocessing, deeper towing may become advantageous because of its improved signal-to-noise ratio, enhanced operational stability, and reduced sensitivity to environmental conditions, while the associated ghost effects can be largely compensated during the proposed preprocessing workflow. This interpretation is further supported by the well-known relation describing the frequencies of free surface ghost notches,

$$f_n = \frac{nV}{2d}, \quad (14)$$

where $(V)$ is the water sound velocity, $(d)$ is the towing depth, and $(n)$ is an integer. In practice, discussions of ghost effects often focus on the observable notches corresponding to $(n = 1,2,3, ...)$, whose frequencies depend on towing depth. However, the same expression formally includes the case $(n = 0)$, indicating a spectral null at zero frequency. The significance of the zero-frequency ghost notch has also been recognised in deterministic deghosting theory (e.g., Amundsen et al., 2013). Free-surface ghosts therefore affect not only discrete notch frequencies but also the lowest-frequency content of the seismic wavefield.

The constructive reuse of the decomposed ghost wavefields is implemented through the stacking operation given by Equation (10). One of the most significant consequences of the proposed framework is the potential recovery of spectral components that would otherwise be discarded together with the ghost wavefield. The field example presented in Figure 11 demonstrates partial recovery of the ghost-related notch near 125 Hz. The synthetic experiments further demonstrate an approximately 6 dB increase in coherent reflection amplitude following ghost reuse, consistent with the constructive superposition described by Equation (10). More importantly, the present formulation suggests that constructive reuse of the decomposed ghost components may also contribute to the recovery of low-frequency energy associated with the zero-frequency null.

Ghost reuse inevitably redistributes residual ghost energy while simultaneously enhancing coherent reflection energy. Consequently, a moderate increase in the background noise level may accompany the coherent signal enhancement. The synthetic experiments suggest that this increase arises primarily from redistribution of residual ghost energy rather than simple amplification of random noise. As summarised in Table 2, the coherent signal gain remains dominant despite the accompanying increase in background noise.

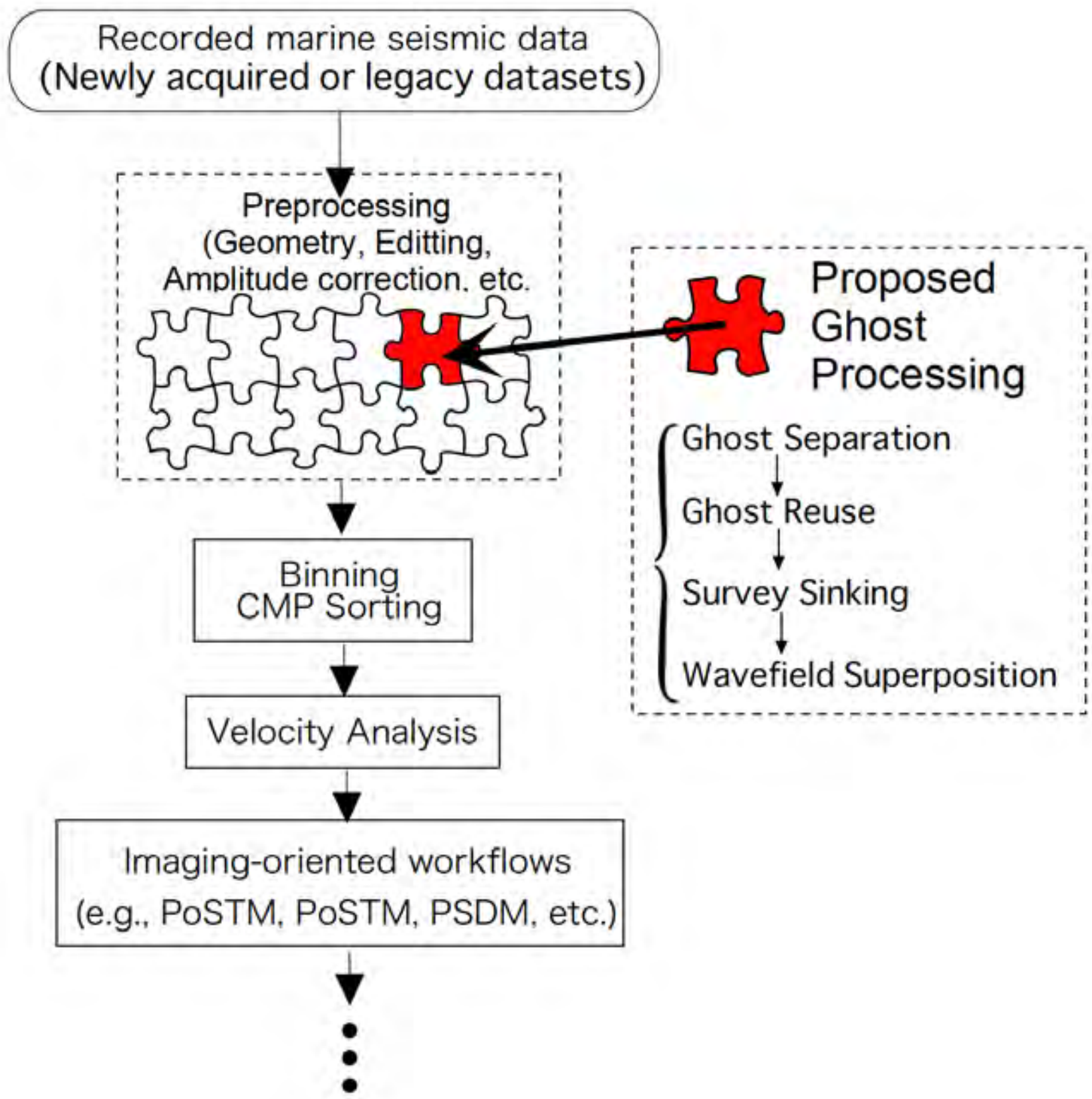

**Figure 12.** Integration of ghost decomposition, survey sinking, wavefield superposition, and constructive reuse of decomposed ghost wavefields into a conventional marine seismic processing workflow. The proposed preprocessing module is incorporated into the conventional workflow before imaging and inversion without modification of the subsequent processing sequence. The framework is therefore intended to complement, rather than replace, established marine seismic data processing workflows.

Figure 12 illustrates the proposed preprocessing workflow, in which ghost decomposition, survey sinking, wavefield superposition, and constructive reuse are incorporated into conventional marine seismic data processing. This workflow allows the proposed framework to be integrated into conventional processing sequences while leaving subsequent imaging and inversion algorithms unchanged. The present study does not suggest replacing existing acquisition strategies. Rather, it indicates that future survey design may benefit from considering acquisition geometry and ghost processing as a unified system rather than as independent stages. From this viewpoint, towing depth becomes a parameter to be optimised jointly with the proposed preprocessing workflow.

An additional implication of the proposed framework concerns the environmental aspects of marine seismic acquisition. Conventional shallow-towed airgun surveys are largely motivated by the need to minimise ghost effects through acquisition geometry. If free surface ghosts can instead be decomposed and constructively reused during processing, deeper towing may become a practical alternative. Besides improving operational stability and reducing sea-state sensitivity, deeper source deployment may also decrease acoustic exposure near the sea surface.

Furthermore, the improved signal quality achieved through ghost reuse may provide greater flexibility in future acquisition design. If comparable imaging performance can be maintained through enhanced processing, it may become possible to reduce the required source energy while preserving image quality. Such a reduction would not only improve acquisition efficiency but also decrease the overall acoustic footprint of marine seismic surveys, potentially reducing acoustic exposure of marine mammals.

Although quantitative evaluation of these environmental benefits lies beyond the scope of the present study, the proposed framework suggests that future marine seismic acquisition may be optimised by jointly considering acquisition geometry, source design, and preprocessing-based ghost processing strategies, thereby balancing imaging performance, operational efficiency, and environmental considerations.

### 5.3 Implications for waveform-based imaging and inversion

Full-waveform inversion (FWI) is particularly sensitive to the availability of low-frequency information because the long-wavelength components of the wavefield play a critical role in mitigating cycle-skipping and constructing reliable initial velocity models (Virieux and Operto, 2009). For this reason, considerable effort has recently been devoted to the development of broadband marine acquisition technologies.

Recent developments include ultra-low-frequency seismic sources (Dellinger et al., 2016; Ronen and Chelminski, 2017; Tellier et al., 2021), variable-depth streamers (Soubaras and Dowle, 2010), over/under streamer configurations (Moldoveanu et al., 2007), and dual-sensor streamer technologies (Itenburg et al., 2006). These acquisition-based approaches seek to mitigate the impact of free surface ghosts and extend the usable bandwidth of marine seismic data.

In contrast, the present framework approaches the same objective from a processing perspective. The field example presented in Figure 11 was acquired using a conventional shallow-towed airgun source and streamer rather than dedicated broadband acquisition technology. Nevertheless, partial recovery of ghost-related spectral notches and associated low-frequency enhancement were observed through the proposed propagation-based preprocessing framework.

Because the proposed framework operates entirely during preprocessing, it can be applied not only to future acquisitions but also to extensive legacy marine seismic datasets acquired before the widespread adoption of broadband acquisition technologies. This capability provides an opportunity to recover additional bandwidth from legacy marine seismic datasets without requiring re-acquisition or modification of subsequent imaging and inversion workflows. The approach is particularly attractive for legacy datasets, for which re-acquisition is often impractical or economically prohibitive. Although the present study does not evaluate the impact of the proposed framework on FWI convergence or inversion accuracy, the observed recovery of low-frequency energy suggests that the proposed framework may provide a useful preprocessing step for future waveform-based imaging and inversion workflows.

### 5.4 Future perspectives

The present study has focused on the fundamental properties of free surface ghost decomposition and constructive reuse within a propagation-based framework. Although the examples considered here are limited to conventional marine acquisition geometries, the underlying formulation is not restricted to two-dimensional datasets or shallow-towed surveys. Potential applications include variable-depth streamer surveys, over/under streamer configurations, dual-sensor streamer technologies, and future broadband acquisition systems.

Although recursive predictive decomposition was adopted in the present study to demonstrate the proposed framework, it is not regarded as the only possible approach to ghost decomposition. The central objective of this study is to demonstrate the constructive reuse of decomposed ghost wavefields. Consequently, any decomposition method capable of providing physically consistent ghost wavefields may be incorporated into the proposed framework.

Interestingly, the multidimensional implementation of recursive predictive decomposition was found to remain stable even at an input S/N ratio of 20 dB. Unlike the one-dimensional delay operator, which propagates the wavefield without geometrical amplitude decay, multidimensional propagation redistributes wavefield energy spatially and introduces geometrical spreading in the reconstructed wavefield. This effect may progressively reduce the local contribution of higher-order propagated components and thereby provide a natural stabilising mechanism for recursive decomposition. A quantitative investigation of this effect remains a subject for future study.

Further development of the multidimensional formulation may also incorporate more realistic source characteristics. The present formulation may provide a useful foundation for future integration with waveform-based imaging and inversion, as well as physics-informed AI-assisted seismic processing. In particular, the explicit decomposition of real and mirrored wavefield components offers a physically interpretable representation that may facilitate the development of data-driven approaches for wavefield classification, reconstruction, and signal enhancement.

Beyond these extensions of the propagation framework, the present formulation is not restricted to an ideal omnidirectional source. Practical effects associated with airgun-array directivity may be incorporated into the propagation operator through an angle-dependent source response, without altering the underlying propagation-based interpretation.

The mirror-source interpretation also suggests the possibility of constructing virtual acquisition geometries through processing alone. Such an approach may provide a conceptual basis for survey sinking strategies, in which deeper effective source or receiver positions are synthesised without modifying the original acquisition system. More generally, this interpretation suggests that marine seismic acquisition and processing need not be regarded as independent stages of the exploration workflow. Instead, they may be viewed as complementary components of a unified wavefield system, in which acquisition geometry and wavefield processing are jointly optimised to maximise the information content of marine seismic observations. Future developments may therefore involve the joint optimisation of acquisition geometry, source design, and ghost processing, rather than treating acquisition and processing as independent stages of the seismic workflow.

## 6. Conclusions

This study has presented a propagation-based framework for the decomposition and constructive reuse of free surface ghost wavefields in marine seismic data. Starting from the finite inverse expansion of the ghost operator, the free surface ghost response was interpreted as a deterministic propagation process rather than solely as an inverse-filtering problem. The formulation was extended from a one-dimensional representation to a multidimensional wavefield framework through the introduction of propagation operators and reciprocity, providing a unified interpretation of source- and receiver-side ghost phenomena.

The multidimensional formulation naturally leads to a decomposition of the recorded wavefield into physically meaningful components associated with real and mirrored sources and receivers. Synthetic and field-data examples demonstrated that the decomposed ghost wavefields can be constructively reused to enhance coherent reflection energy, improve reflector continuity, and partially restore spectral amplitudes degraded by the free surface ghost.

The proposed framework may also contribute to the recovery of low-frequency energy attenuated by the free surface ghost response. As an entirely processing-based approach, it is applicable to both future broadband acquisitions and legacy marine seismic datasets. Although not evaluated here, the recovery of low-frequency information may also benefit waveform-based imaging and inversion.

More broadly, this study demonstrates that free surface ghosts can be interpreted not solely as unwanted noise, but as reusable physical wavefield components for coherent signal enhancement. This perspective may contribute to the joint optimisation of acquisition geometry, source design, and ghost processing in future marine seismic exploration.